\documentclass[a4paper,11pt]{article}
\usepackage{amsmath}
\usepackage{graphicx}
\usepackage{amssymb}
\usepackage{cite}
\usepackage{natbib}
\usepackage[hidelinks]{hyperref}

\hypersetup{
  colorlinks = true,     
  urlcolor = blue,       
  linkcolor = blue,      
  citecolor = red         
}

\newcommand{\rmi}{\mathrm{i}}
\newcommand{\rme}{\mathrm{e}}

\providecommand{\keywords}[1]
{
  \small	
  \textbf{\textit{Keywords---}} #1
}

\begin{document}

\title{The Requirement of (at least) Complex Structure for Quantum Mechanics}

\author{M.P. Vaughan\footnote{\emph{Corresponding author}: martin.vaughan@physics.org}}

\date{}

\maketitle

\begin{abstract}
It is argued that many real-valued constructions of quantum mechanics are only \emph{nominally} real in that the operators and states are restricted to impose a \emph{complex structure} on the Hilbert space. That is, complex algebra between pairs of elements representing complex numbers is preserved in these formulations. It is therefore mistaken to think of these constructions as being `real' as they are actually a representation of complex linear algebra. It is then shown that under the assumptions of state normalisation and strict energy conservation, this complex structure (at the very least) is required to allow for time evolution in quantum mechanics. This is only a \emph{minimum} requirement, as our arguments do not preclude the formulation of quantum mechanics in terms of hyper-complex entities, such as quaternions.
\end{abstract}

\keywords{complex structure, complex quantum mechanics, real quantum mechanics, time evolution}


\section{Introduction}
A perennial question in the topic of quantum foundations concerns the role of complex numbers in the formulation of quantum theory. This has lead many researchers to question whether complex numbers are truly necessary or whether there is not some equally viable alternative framework based only on real number algebra. This has resulted in many endeavours to construct formulations of quantum mechanics based on real Hilbert spaces~\cite{stueckelberg1960quantum, myrheim1999quantum, mckague2009simulating, aleksandrova2013real, singh2026quantum, hita2026quantum}. 

Although often cited as the seminal work for such formulations, Stueckelberg's 1960 paper~\cite{stueckelberg1960quantum} was actually  based on lectures intended to explain to students the \emph{necessity} of the imaginary unit `$\rmi$' in quantum theory.  Stueckelberg's central argument was that any valid formulation of quantum theory should retain the uncertainty principle, which in turn, required the non-vanishing of the commutator between conjugate pairs of physical observables. 

Stuckelberg then argued that this was only possible through the introduction of a universal antisymmetric operator $\tilde{J}$ with exactly the same properties as `$\rmi$' and that commuted with all other physical observables. Aleksandrova \emph{et al}~\cite{aleksandrova2013real} referred to this requirement as \emph{Stueckelberg's rule}. In their real Hilbert space formulation they proposed that the function of this operator was provided by a \emph{universal quantum bit}, which they interpreted physically.

Then, in landmark 2021 Nature publication, Renou \emph{et al}~\cite{renou2021quantum} demonstrated that real-valued quantum mechanics could be experimentally falsified. These researchers analysed the possible measurements made on entangled states distributed amongst several agents using both real and complex quantum mechanics, showing that the two formulations yielded different predictions. Later experimental tests of these predictions by Chen \emph{et al}~\cite{chen2022ruling} confirmed the complex version of the theory.

Whilst this may have appeared to be a conclusive confirmation of the necessity for complex numbers, Hita \emph{et al}~\cite{hita2026quantum} have recently challenged this conclusion, arguing that Renou \emph{et al}'s assumptions were too restrictive and introducing a real-valued formulation of quantum mechanics that replicates the predictions of the complex theory for all multipartite experiments. Further support for this position came from an independently developed real-valued formulation by  Hoffreumon and Woods~\cite{hoffreumon2025quantum}.

Now, it is the contention of this paper that certain constructions of quantum mechanics based on a real Hilbert space (including many of those discussed above) are only \emph{nominally} real in that they preserve a \emph{complex structure}. That is, the complex arithmetic between the operators and vectors of the corresponding complex Hilbert space is still retained. The complex numbers themselves are only disguised as real numbers due to a convention of \emph{representation}. Specifically, we shall argue that these complex elements still exist in the nominally real Hilbert space in their $2\times2$ (real) matrix forms. 

It is then shown in Section~\ref{sec:complex-structure} that, given the requirements of state normalisation and strict energy conservation, any real Hilbert space construction allowing the time evolution of a quantum state must have \emph{at least} this complex structure. By `at least', we mean that more complicated structures supporting, for example, quaternions~\cite{hamilton1840new}, may also satisfy the requirements for time evolution.

We argue that the claim that such nominally real representations have done away with the need for complex numbers is mistaken and misleading. Moreover, if the purpose of research into such real representations is to gain deeper insight into the fundamental nature of quantum mechanics, then this belief is counter-productive. What we should truly be recognising is that complex or hyper-complex algebra appears to be necessary for describing quantum mechanics but that such mathematics may be supported (as is indeed well-known to pure mathematicians) via the use of real matrices. 

\section{Nominally real representations}\label{sec:nominally-real}
\subsection{Representing complex numbers}
A defining characteristic of the real numbers is that they are \emph{one-dimensional}. Geometrically speaking, this means that we can map every element of $\mathbb{R}$ to a unique point on an infinitely long one-dimensional line. This has the consequence that the reals numbers are inherently \emph{ordered}. That is, for any two distinct real numbers $x$ and $y$, we can say that one is greater than the other. We can describe this by saying that the real numbers constitute an \emph{ordered field}. This is the characteristic that makes real numbers so useful for modelling physical observables. 

A complex number is in essence a \emph{two-dimensional} number requiring \emph{two} ordered fields for its specification. The \emph{notation} that we use to represent it is entirely a matter of preference. In terms of what we might call the `scalar' representation, we have a number of choices. We might denote a complex number $z$ in terms of real numbers $a$ and $b$ using the imaginary unit `$\rmi$' as $z = a + \rmi b$ or in polar form in terms of real numbers $r$ and $\theta$ as $z = r\rme^{\rmi\theta}$. We might also dispense with the symbol `$\rmi$' and, following Hamilton's lead, write a complex number as an algebraic couple $z = (a, b)$~\cite{hamilton1831theory}, along with the appropriate definitions for arithmetic. 

Alternatively, we might employ the well-known isomorphism between the scalar form of a complex number and its $2\times2$ matrix representation

\begin{align}
a + \rmi b &\cong \left[\begin{array}{cc}
a & -b \\
b & a\end{array}\right]. \label{eq:isomorphism}
\end{align}

\noindent The relation to the scalar form may then be drawn out by writing the matrix in the form

\begin{align}
\left[\begin{array}{cc}
a & -b \\
b & a\end{array}\right] &= a\hat{I}_{2} + b\hat{J}_{2}, \label{eq:2by2asIandJ}
\end{align}

\noindent where

\begin{align}
\hat{I}_{2} \equiv \left[\begin{array}{cc}
1 & 0 \\
0 & 1\end{array}\right]~\mathrm{and}~ \hat{J}_{2} \equiv \left[\begin{array}{cc}
0 & -1 \\
1 & 0\end{array}\right]. \label{eq:IandJdef}
\end{align}

\noindent Here we see that, since $\hat{J}_{2}^{2} = -\hat{I}$, the $\hat{J}_{2}$ matrix plays the same role as the imaginary unit `$\rmi$' in the scalar representation. An important point to note (which the reader may easily prove for themselves) is that a $2\times2$ matrix will have the form of \eqref{eq:isomorphism} \emph{if and only if} it commutes with $\hat{J}_{2}$.

One advantage of the matrix representation is that, using Euler's identity to express the complex number in polar form, it clearly illustrates the nature of complex multiplication.

\begin{align}
r\rme^{\rmi\theta} &\cong r\left[\begin{array}{cc}
\cos\theta & -\sin\theta \\
\sin\theta & \cos\theta\end{array}\right]. \label{eq:isomorphism-polar}
\end{align}

\noindent Here it is easy to see that multiplication then involves both a \emph{scaling} by the positive real number $r$ and a \emph{rotation} by the phase $\theta$. It is interesting to note that, whilst real number multiplication usually only involves a scaling, the exception is multiplication by $-1$, which can be viewed as a rotation by $\pi$. 

Note that despite the fact that the matrices in \eqref{eq:isomorphism} and \eqref{eq:isomorphism-polar} have only real elements \emph{they still represent complex numbers}. They are simply an alternative representation using two real numbers to specify them. The fact that complex algebra can be described in terms of these two real numbers should not be taken as implying it is in some fundamental way `real'. \emph{Real} algebra can be described in terms of a \emph{single} ordered field, which is not possible for the complex numbers. The defining feature of complex numbers is that they are two dimensional.

\subsection{Operators on a Hilbert space}
\subsubsection{Constructing nominally real operators}
Let us now consider how we may construct \emph{nominally real} operators from the operators on an $N$-dimensional complex Hilbert space $\mathbb{C}^{N}$. The operators on $\mathbb{C}^{N}$ may be represented by $N\times N$ matrices containing complex elements. We shall adopt the simple tactic of replacing each such element by its $2\times2$ matrix equivalent, thereby producing a $2N\times 2N$ matrix containing only real elements.

For any given complex matrix $\hat{A}$, we would then have the mapping (with a slight abuse of notation for illustrative purposes)

\begin{align}
\left[\begin{array}{ccc}
A_{00} & A_{01} & \ldots \\
A_{10} & A_{11} & \ldots \\
\vdots & \vdots & \ddots\end{array}\right] &\to \left[\begin{array}{ccc}
\left[\begin{array}{cc} A_{00}^{r} & -A_{00}^{i} \\ A_{00}^{i} & A_{00}^{r}\end{array}\right] & \left[\begin{array}{cc} A_{01}^{r} & -A_{01}^{i} \\ A_{01}^{i} & A_{01}^{r}\end{array}\right] & \ldots \\ \\
\left[\begin{array}{cc} A_{10}^{r} & -A_{10}^{i} \\ A_{10}^{i} & A_{10}^{r}\end{array}\right] & \left[\begin{array}{cc} A_{11}^{r} & -A_{11}^{i} \\ A_{11}^{i} & A_{11}^{r}\end{array}\right] & \ldots \\
\vdots & \vdots & \ddots\end{array}\right], \label{eq:mapping}
\end{align}

\noindent where $A_{ij}^{r}$ and $A_{ij}^{i}$ denote the real and imaginary components respectively of the $A_{ij}$ element. Denoting this new $2N\times2N$ matrix by $\tilde{A}$, it is clear that this can be written as

\begin{align}
\tilde{A} &= \hat{A}^{r}\otimes\hat{I}_{2} + \hat{A}^{i}\otimes\hat{J}_{2}, \label{eq:A-tilde}
\end{align}

\noindent where `$\otimes$' denotes the Kronecker product and the matrices $\hat{A}^{r}$ and $\hat{A}^{i}$ are the real and imaginary components the $\hat{A}$ matrix.

At this point, it is important to appreciate that the complex arithmetic of the elements of $\hat{A}$ is still preserved when using `real' matrices constructed in this way. For addition, this is easy to see. For multiplication, we can use the mixed-product property of the Kronecker products to show that, for another real matrix $\tilde{B}$ constructed from its complex equivalent via \eqref{eq:mapping}, we would have

\begin{align}
\tilde{A}\tilde{B} &= \left(\hat{A}^{r}\hat{B}^{r} - \hat{A}^{i}\hat{B}^{i}\right)\otimes\hat{I}_{2} + \left(\hat{A}^{r}\hat{B}^{i} + \hat{A}^{i}\hat{B}^{r}\right)\otimes\hat{J}_{2}, \nonumber \\ 
&= (\hat{A}\hat{B})^{r}\otimes\hat{I}_{2} + (\hat{A}\hat{B})^{i}\otimes\hat{J}_{2}. \nonumber
\end{align}

\noindent Here one recognizes the general form of complex multiplication and we emphasise this by asserting that the $\tilde{A}$ and  $\tilde{B}$ matrices retain what we shall call a \emph{complex structure}.

\subsubsection{Permutations of the Kronecker product}
Whilst we have used the `$\otimes$' to denote the Kronecker product, it is more often used in the literature to denote a general \emph{tensor product}, especially in the context of constructing states and operators for bipartite systems. In these cases, the order of the operands makes no \emph{physical} difference. When using the Kronecker product to achieve this, though, it is generally the case that $\hat{A}\otimes\hat{B} \ne \hat{B}\otimes\hat{A}$. However, such products \emph{are} related in that one is a permutation of the other. That is, we can always find a real permutation matrix $\hat{P}$ such that 

\begin{align}
\hat{A}\otimes\hat{B} &= \hat{P}(\hat{B}\otimes\hat{A})\hat{P}^{T}, \nonumber
\end{align} 

\noindent where the $T$ superscript denotes transposition. Such permutations just change the row and column orderings so have no physical significance for the entities the matrices represent. 

Now, it can be shown that the \emph{same} $2N\times2N$ permutation matrix $\tilde{P}$ that takes the product $\hat{A}_{N}\otimes\hat{I}_{2}$ to $\hat{I}_{2}\otimes\hat{A}_{N}$ (where $\hat{A}_{N}$ is an $N\times N$ matrix) \emph{also} takes $\hat{A}_{N}\otimes\hat{J}_{2}$ to $\hat{J}_{2}\otimes\hat{A}_{N}$. That is, for matrices of the form of \eqref{eq:A-tilde}, we can always find a permutation matrix $\tilde{P}$ such that 

\begin{align}
\tilde{P}\tilde{A}\tilde{P}^{T} &= \tilde{P}(\hat{A}^{r}\otimes\hat{I}_{2} + \hat{A}^{i}\otimes\hat{J}_{2})\tilde{P}^{T}, \nonumber \\
&= \hat{I}_{2}\otimes\hat{A}^{r} + \hat{J}_{2}\otimes\hat{A}^{i} \equiv \tilde{A}'. \label{eq:permutation}
\end{align}

\noindent This means that $\tilde{A}'$ and $\tilde{A}$ are equivalent representations of the same operator and we may always transform from one form to the other by applying this permutation universally to all the expressions we are using. 

Since we may always transform back via the inverse permutation operation, this means the all the mathematical properties of the matrices constructed using \eqref{eq:mapping} are retained. In other words, matrices of the form of \eqref{eq:permutation} \emph{also} have a complex structure. The only difference is that such matrices will have the manifest form

\begin{align}
\tilde{A}' &= \left[\begin{array}{cc}
\hat{A}^{r} & -\hat{A}^{i} \\
\hat{A}^{i} & \hat{A}^{r}\end{array}\right]. \label{eq:permutation-manifest}
\end{align}

\noindent More generally, we assert that \emph{any} permutation carried out universally on operators of the form of \eqref{eq:mapping} yields a set of operators with complex structure.

It is salient to note that matrix operators of the form of either \eqref{eq:A-tilde} or \eqref{eq:permutation} are often given in the literature as examples of `real' operators along with the claim that complex numbers have been removed from them. For example, Eq. (4) of Ref~\cite{hita2026quantum} or Eq. (4) of Ref.~\cite{hoffreumon2025quantum}. We insist that \emph{no such removal has occurred}. These are just \emph{different representations} of complex operators. 

Another important point to note is that whilst matrices such as \eqref{eq:A-tilde} or \eqref{eq:permutation} belong to the space of operators mapping vectors of a $2N$ dimensional real Hilbert space to themselves (the \emph{endomorphism} of $\mathbb{R}^{2N}$), they strictly belong to a \emph{proper subset} of such operators. For example, we do not include in the set of allowed operators matrices of the form $\hat{A}\otimes\sigma_{x}$ or $\hat{A}\otimes\sigma_{z}$ (where $\sigma_{x}$ and $\sigma_{z}$ are Pauli matrices). It is this restriction to a proper subset (or subspace) of the endomorphism of $\mathbb{R}^{2N}$ that imposes the complex structure on the representation.

\section{The requirement of (at least) complex structure}\label{sec:complex-structure}
\subsection{Time evolution of a state}
Let us now consider a real Hilbert space $\mathcal{H}$ of finite dimension without making any assumptions about whether it has a complex structure or not. In accordance with the usual postulates of quantum mechanics, we shall assume that a physical system may be represented by a unit vector $|\Psi\rangle$ of the Hilbert space such that

\begin{align}
\langle\Psi|\Psi\rangle &= 1, \label{eq:normalised}
\end{align}

\noindent  where, being real, $\langle\Psi|$ is the \emph{transpose} of $|\Psi\rangle$. (The generalisation to density matrices to represent a state is analysed in Appendix~\ref{app:density-matrices}).

To first order, the time evolution of the state may be given as

\begin{align}
|\Psi(t+\delta t)\rangle &= (\tilde{I} + \tilde{\Omega}\delta t)|\Psi(t)\rangle, \label{eq:time-increment}
\end{align}

\noindent where $\tilde{I}$ is the identity matrix on $\mathcal{H}$ and $\tilde{\Omega}$ is some real operator to be determined. Rearranging \eqref{eq:time-increment} and taking the limit $\delta t \to 0$ we readily find

\begin{align}
\frac{d}{dt}|\Psi(t)\rangle &= \tilde{\Omega}|\Psi(t)\rangle. \label{eq:time-derivative}
\end{align}

Now, in order to satisfy the normalisation condition of \eqref{eq:normalised}, we must have

\begin{align}
\langle\Psi(t+\delta t)|\Psi(t+\delta t)\rangle &= \langle\Psi(t)|(\tilde{I} + \tilde{\Omega}^{T}\delta t)(\tilde{I} + \tilde{\Omega}\delta t)|\Psi(t)\rangle = 1, \nonumber
\end{align}

\noindent which implies

\begin{align}
\tilde{\Omega}^{T} &= -\tilde{\Omega}. \label{eq:antisymmetric}
\end{align}

\noindent In other words, $\tilde{\Omega}$ must be an \emph{antisymmetric} matrix.

Solving \eqref{eq:time-derivative}, we find that the general time dependence of a state is given by

\begin{align}
|\Psi(t)\rangle &= \tilde{U}(t)|\Psi(0)\rangle, \label{eq:time-dependence}
\end{align}

\noindent where

\begin{align}
\tilde{U}(t) &= \rme^{\tilde{\Omega}t}. \label{eq:evolution-op}
\end{align}

\noindent Note that, due to the antisymmetry of $\tilde{\Omega}$, $\tilde{U}(t)$ is a member of the \emph{special orthogonal group} on the space  of operators on $\mathcal{H}$~\cite{vaughan2026time}. That is, this is a real orthogonal matrix with a determinant of $+1$.

In practice, when constructing a real Hilbert space via the tacit use of \eqref{eq:mapping}, the use of unit vectors of the resultant $2N$ dimensional real space to model physical states can become problematic. This is due to the fact that column vectors of the complex Hilbert space will be mapped to $2N\times2$ \emph{matrices} of the real space. As a result, it is more common to represent a state by a \emph{density matrix} (which still needs a modifying factor of $1/2$ to keep the mathematics straight). However, the analysis we follow here will still apply to density matrices, as described in Appendix~\ref{app:density-matrices}.

\subsection{Energy conservation}\label{sec:energy-conservation}
In complex quantum mechanics, physical observables are modelled by \emph{Hermitian} operators. In this case, the real part of a Hermitian matrix will be \emph{symmetric}, whilst the imaginary part will be \emph{antisymmetric}. Hence a real matrix of the form of \eqref{eq:A-tilde} (or \eqref{eq:permutation}) representing a Hermitian matrix will be symmetric and will have real eigenvalues, as required.

We shall construct the Hamiltonian operator formally in its diagonal representation as 

\begin{align}
\tilde{H} &= \sum_{k} E_{k}|\phi_{k}\rangle\langle\phi_{k}|, \label{eq:Hamiltonian}
\end{align}

\noindent for energy eigenvalues $E_{k}$ and real eigenvectors $|\phi_{k}\rangle$. This shall be taken to be the Hamiltonian for the total system and, in line with the conservation of energy, we shall take it to have no time dependence. More specifically, we shall assert that the energy expectation value for any state does not change with time. That is,

\begin{align}
\frac{d}{dt}\langle E\rangle &= \frac{d}{dt}\langle\Psi|\tilde{H}|\Psi\rangle = 0, \label{eq:energy-conservation}
\end{align}

\noindent for any state $|\Psi\rangle$ of the \emph{total} system (in a multipartite system, energy might be transferred from one subsystem to another but the \emph{total} energy must be conserved).

Using \eqref{eq:time-derivative} and \eqref{eq:antisymmetric}, the time-derivative of $\langle E\rangle$ becomes

\begin{align}
\frac{d}{dt}\langle\Psi|\tilde{H}|\Psi\rangle &= \langle\Psi|(\tilde{H}\tilde{\Omega} - \tilde{\Omega}\tilde{H})|\Psi\rangle = 0. \nonumber
\end{align}

\noindent For this to  hold for arbitrary $|\Psi\rangle$, we must have

\begin{align}
\tilde{H}\tilde{\Omega} - \tilde{\Omega}\tilde{H} &= 0. \label{eq:HOmega-condition}
\end{align}

\noindent (In Appendix~\ref{app:density-matrices} we show that the same condition emerges when analysing the time dependence of a density matrix). In the energy representation, we then have

\begin{align}
\langle\phi_{i}|(\tilde{H}\tilde{\Omega} - \tilde{\Omega}\tilde{H})|\phi_{j}\rangle &=  \sum_{k} (H_{ik}\Omega_{kj} - \Omega_{ik}H_{kj}), \nonumber \\
&= \sum_{k} (\delta_{ik}E_{k}\Omega_{kj} - \Omega_{ik}E_{k}\delta_{kj}), \nonumber \\
&= (E_{i} - E_{j})\Omega_{ij} = 0. \label{eq:HOmega-commute}
\end{align}

\noindent This can only hold generally if for every non-zero element $\Omega_{ij}$ we have $E_{i} = E_{j}$. The bare minimum for this condition to be met is that each energy eigenvalue is 2-fold degenerate. In this particular scenario we must also have that the dimensionality of $\mathcal{H}$ is even (i.e. $\mathcal{H} = \mathbb{R}^{2N}$).

In this case, the energy eigenvalues can be arranged in pairs with the corresponding elements of $\tilde{\Omega}$ sitting in the off-diagonal positions of $2\times2$ matrices arranged into a block diagonal matrix. Note also that, since $\tilde{\Omega}$ is antisymmetric, the $\Omega_{ij}$ elements in each of these $2\times2$ blocks must be equal.

The $2N\times2N$ $\tilde{H}$ and $\tilde{\Omega}$ matrices will then have the following forms:

\begin{align}
\tilde{H} &= \left[\begin{array}{ccc}
\left[\begin{array}{cc} E_{0} & 0 \\ 0 & E_{0}\end{array}\right] & \left[\begin{array}{cc} 0 & 0 \\ 0 & 0\end{array}\right] & \ldots \\ \\
\left[\begin{array}{cc} 0 & 0 \\ 0 & 0\end{array}\right] & \left[\begin{array}{cc} E_{1} & 0 \\ 0 & E_{1}\end{array}\right] & \ldots \\
\vdots & \vdots & \ddots\end{array}\right], \nonumber \\
&= \left[\begin{array}{ccc}
E_{0} & 0 & \ldots \\ \\
0 & E_{1} & \ldots \\
\vdots & \vdots & \ddots\end{array}\right]\otimes\hat{I}_{2} \label{eq:hamiltonian-matrix}
\end{align}

\noindent and

\begin{align}
\tilde{\Omega} &= \left[\begin{array}{ccc}
\left[\begin{array}{cc} 0 & -\Omega_{0} \\ \Omega_{0} & 0\end{array}\right] & \left[\begin{array}{cc} 0 & 0 \\ 0 & 0\end{array}\right] & \ldots \\ \\
\left[\begin{array}{cc} 0 & 0 \\ 0 & 0\end{array}\right] & \left[\begin{array}{cc} 0 & -\Omega_{1} \\ \Omega_{1} & 0\end{array}\right] & \ldots \\
\vdots & \vdots & \ddots\end{array}\right], \nonumber \\
&= \left[\begin{array}{ccc}
\Omega_{0} & 0 & \ldots \\ \\
0 & \Omega_{1} & \ldots \\
\vdots & \vdots & \ddots\end{array}\right]\otimes\hat{J}_{2}. \label{eq:omega-matrix}
\end{align}

\noindent Clearly, these matrices commute with one another, so satisfying \eqref{eq:HOmega-condition}. It is also clear that these matrices \emph{have the complex structure} described in Section \ref{sec:nominally-real}. 

In passing, we note that by putting $\Omega_{i} = -E_{i}/\hbar$ (so giving the $\Omega_{i}$ dimensions of angular frequency), we can obtain the general form of the Schr{\"o}dinger equation. The Hamiltonian would now be given by

\begin{align}
\tilde{H} &= \hbar\tilde{J}\tilde{\Omega}, \nonumber
\end{align}

\noindent where 

\begin{align}
\tilde{J} &\equiv \hat{I}_{N}\otimes\hat{J}_{2} \nonumber
\end{align}

\noindent and $\hat{I}_{N}$ is the $N\times N$ identity matrix. Operating on a general state and using \eqref{eq:time-derivative} we then have

\begin{align}
\tilde{H}|\Psi(t)\rangle &= \hbar\tilde{J}\frac{d}{dt}|\Psi(t)\rangle. \label{eq:schrodinger}
\end{align}

\noindent Note that $\tilde{J}^{2} = -\tilde{I}$, where $\tilde{I}$ is the identity matrix on $\mathbb{R}^{2N}$, so functions as the imaginary unit. Meanwhile $\tilde{J}\tilde{\Omega}$ is now a symmetric matrix, matching $\tilde{H}$. The fact that we have a total time derivative rather than a partial derivative is merely due to an easily removed tacit assumption that $|\Psi(t)\rangle$ depends only on $t$.

Although arrived at by an arbitrary assumption, \eqref{eq:schrodinger} now shares with the Schr{\"o}dinger equation an encapsulation of the fundamental physical concept that \emph{energy is the generator of temporal change}.

\subsection{Beyond complex structure}
It should be immediately conceded that, whilst we have argued that we require \emph{at least} complex structure for time evolution, there is no obvious reason why we cannot go beyond this. For example, instead of the energy eigenvalues being only 2-fold degenerate, \eqref{eq:HOmega-condition} would also be satisfied if $\tilde{H}$ and $\tilde{\Omega}$ were composed of $4\times4$ blocks of the form

\begin{align}
\left[\begin{array}{cccc}
E_{i} & 0 & 0 & 0 \\
0 & E_{i} & 0 & 0 \\
0 & 0 & E_{i} & 0 \\
0 & 0 & 0 & E_{i}\end{array}\right]~\mathrm{and}~\left[\begin{array}{cccc}
0 & -\Omega_{i1} &  -\Omega_{i2} &  -\Omega_{i3} \\
\Omega_{i1} &0 &  -\Omega_{i3} &  \Omega_{i2} \\
\Omega_{i2} & \Omega_{i3} & 0 &  -\Omega_{i1} \\
\Omega_{i3} & -\Omega_{i2} & \Omega_{i1} & 0\end{array}\right]. \nonumber
\end{align}

Whilst, on first glance, this example may seem arbitrary, it has been chosen since the form of these matrices provide a representation of \emph{quaternions} (the diagonal matrix on the left being a real part and the antisymmetric matrix on the right representing a pure quaternion of the form $\Omega_{i1}\rmi + \Omega_{i2}\mathrm{j} + \Omega_{i3}\mathrm{k}$). Note, for example, that (denoting the right-hand matrix by $\hat{J}_{4}$) if we normalise the elements so that $\Omega_{i1}^{2}+ \Omega_{i2}^{2} + \Omega_{i3}^{2} = 1$, then $\hat{J}_{4}^{2} = -\hat{I}_{4}$, where $\hat{I}_{4}$ is the $4\times4$ identity matrix.

\section{Discussion and conclusions}
It is certainly not the case that \emph{all} proposed real-valued formulations of quantum mechanics have complex structure. In particular, the 2-dimensional Hilbert space of a \emph{rebit}, a real-valued version of a qubit introduced by Caves \emph{et al}~\cite{caves2001entanglement}, lacks such structure. In this case, however, the authors were not attempting to provide an equivalent formulation of quantum mechanics but rather to propose a foil theory to compare and contrast with the complex theory in order to better understand the phenomenon of entanglement in multipartite systems. Indeed, they showed that the two approaches \emph{do} give different predictions for the separability of multipartite systems. This result was used by Renou \emph{et al}~\cite{renou2021quantum} in their argument that real and complex-valued formulations give different predictions for measurements on entangled resources shared between three agents. 

Consider the endomorphism of $\mathbb{R}^{2}$ in which we would represent a single rebit. This is spanned by just four real matrices: the symmetric matrices $\hat{I}_{2}$, $\sigma_{x}$ and $\sigma_{z}$, and the antisymmetric matrix $\rmi\sigma_{y} = \hat{J}_{2}$. Now, we could \emph{impose} a complex structure on this space by disallowing $\sigma_{x}$ and $\sigma_{z}$, although this would be highly restrictive. In particular, we could not represent many of the more interesting quantum gates, including the Hadamard gate, given by $H = (\sigma_{x} + \sigma_{z})/\sqrt{2}$. (The role of complex numbers in the time evolution of quantum gates is discussed in more detail in Ref.~\cite{vaughan2026time}). Moreover, as our earlier arguments have shown, in order to have any time evolution, such a rebit must have degenerate energy levels.

So, in order to properly reproduce the dynamics of a \emph{qubit} on a real Hilbert space, $\mathbb{R}^{2}$ will certainly \emph{not} do. We need to double the dimensions of the original space $\mathbb{C}^{2}$ and use $\mathbb{R}^{4}$. However, so far as the space of operators goes, we shall only be using a \emph{subspace} of the endomorphism of $\mathbb{R}^{4}$ that imposes the complex structure.

It is when we come to model multipartite systems that the construction of real-valued formulations becomes potentially problematic. For two systems represented on complex Hilbert spaces $\mathbb{C}^{M}$ and $\mathbb{C}^{N}$, the joint system would be represented on the tensor product of the two spaces, having $MN$ dimensions. Since this is complex, matrices on this space would then have $2(MN)^{2}$ degrees of freedom. To model this using a real Hilbert space, we could apply the mapping of \eqref{eq:mapping} to obtain a space of $2MN$ dimensions. The space of operators will then have dimensions of $4(MN)^{2}$. However, since in we only use \emph{half} of these operators (the Kronecker products of the basis of the original space with $\hat{I}_{2}$ and $\hat{J}_{2}$), the degrees of freedom of are reduced to $2(MN)^{2}$, matching that of the complex space.

Problems arise if we perform the mapping of \eqref{eq:mapping} to the complex spaces $\mathbb{C}^{M}$ and $\mathbb{C}^{N}$ first, to obtain spaces $\mathbb{R}^{2M}$ and $\mathbb{R}^{2N}$, and \emph{then} took the tensor product of the real spaces to obtain a Hilbert space of $4MN$ dimensions. The space of operators will then have dimensions of $16(MN)^{2}$, so that even after only taking half of these we would still have  $8(MN)^{2}$ degrees of freedom. The two processes are therefore not equivalent and we need to think carefully about how to construct our joint Hilbert space when using real numbers. 

It is in the construction of frameworks to model multipartite systems that the different consequences for separability pointed out by Caves \emph{et al}~\cite{caves2001entanglement} creep in. In a real-valued formulation, physical operators and states (in the form of density matrices) are support entirely by symmetric matrices. Now, the symmetric matrices of  a joint system may contain elements formed from the Kronecker products of the antisymmetric elements of the subspaces. Since these cannot come from the independent states of the subsystems, the existence of such components in the density matrix of the joint system would imply that it is entangled.

Let us consider, for example, the endomorphism of a vector space representing a bipartite system of two \emph{rebits}. This will be a space of $4\times4$ matrices for which the real symmetric matrix $(\mathrm{i}\sigma_{y})\otimes(\mathrm{i}\sigma_{y})$ might be a component of a density matrix. However, since $(\mathrm{i}\sigma_{y})$ is antisymmetric,  these components could not have come from the states of the independent rebits, indicating that the joint system is entangled. 

On the other hand, the element $(\mathrm{i}\sigma_{y})\otimes(\mathrm{i}\sigma_{y})$ would naturally show up as a component of the density matrix of a single \emph{qubit} modelled on $\mathbb{R}^{4}$, being the Kronecker product of the imaginary part with $\hat{J}_{2}$, which is clearly \emph{not} an entangled state. We therefore need to be very careful in how we go about constructing a joint system to avoid situations in which the real formulation is indicating entanglement whilst the complex version is pointing to separability.

Avoiding this problem is, in fact, quite simple. We just need to remember that the (successful) real-valued systems we are using have a complex structure imposed, so whenever we have Kronecker products of an element with $\hat{I}_{2}$ or $\hat{J}_{2}$ this is actually equivalent to multiplying these elements by `1' or `$\rmi$'. The easiest way to do this is simply to convert back to scalar representation, take the tensor product of the two complex Hilbert spaces and then convert back to the real representation using \eqref{eq:mapping}.

Alternatively, we could come up with some more convoluted solution, such as introducing some system for flagging the real and imaginary parts and treating the multiplication of these differently to the other matrix elements. The result, however, would be same. We would simply be introducing some mechanism for preserving the complex structure of the Hilbert spaces we were constructing.

We conclude then that, whilst we can use real Hilbert spaces to replicate the results of the complex formulation of quantum mechanics, these spaces have a \emph{complex structure} imposed on them and that we require (at least) this structure to allow for time evolution. These constructions are therefore only \emph{nominally} real in the sense that they are still reproducing complex linear algebra. 

The symbol `$\rmi$' does not define complex numbers - their algebraic properties do and these are always ultimately a function of the \emph{two} ordered fields that specify them. Complex algebra \emph{cannot} be defined in terms of a \emph{single} ordered field. To believe that such approaches actually do away with complex numbers, simply because we only use real numbers to specify them, is to miss a valuable insight into the nature of quantum dynamics.

\bibliographystyle{unsrt}
\bibliography{real-quantum}

\appendix
\section{Appendix}
\subsection{Density matrices}\label{app:density-matrices}
It is a common practice in many areas of applied quantum mechanics to represent a physical state, not by a unit state vector $|\Psi\rangle$ of the Hilbert space $\mathcal{H}$, but by a \emph{density matrix} $\hat{\rho}$ defined via the outer product

\begin{align}
\hat{\rho} &\equiv |\Psi\rangle\langle\Psi|. \label{eq:density-matrix-def}
\end{align}

\noindent Such a density matrix then contains the same information as the state $|\Psi\rangle$ but in matrix form. The normalisation condition $\langle\Psi|\Psi\rangle = 1$ then becomes

\begin{align}
\langle\Psi|\Psi\rangle &= \sum_{i} \langle\Psi|\phi_{i}\rangle\langle\phi_{i}|\Psi\rangle = \sum_{i} \langle\phi_{i}\Psi\rangle\langle\Psi|\phi_{i}\rangle, \nonumber \\
&= \mathrm{Tr}[\hat{\rho}] = 1, \label{eq:density-matrix-norm}
\end{align}

\noindent where $\mathrm{Tr}[\cdot]$ denotes the \emph{trace operator} and the $\{|\phi_{i}\rangle\}$ is some complete, orthonormal basis set such that $\sum_{i} |\phi_{i}\rangle\langle\phi_{i}|$ is the identity operator on $\mathcal{H}$.

In terms of the density matrix, the \emph{expectation value} of an operator $\hat{A}$ on $\mathcal{H}$ is given by

\begin{align}
\langle\hat{A}\rangle &= \langle\Psi|\hat{A}|\Psi\rangle = \sum_{i}\langle\Psi|\hat{A}|\phi_{i}\rangle\langle\phi_{i}|\Psi\rangle, \nonumber \\
&= \sum_{i}\langle\phi_{i}|\Psi\rangle\langle\Psi|\hat{A}|\phi_{i}\rangle, \nonumber \\
&= \mathrm{Tr}[\hat{\rho}\hat{A}]. \label{eq:expectation-rho}
\end{align}

Assuming that $\mathcal{H}$ is a real Hilbert space and denoting the density operator by $\tilde{\rho}$ accordingly, we can use \eqref{eq:time-derivative} to find the time dependency of  $\tilde{\rho}(t)$ as

\begin{align}
\frac{d}{dt}\tilde{\rho} &= \tilde{\Omega}\tilde{\rho} - \tilde{\rho}\tilde{\Omega} = [\tilde{\Omega},\tilde{\rho}], \label{eq:rho-time-dependence}
\end{align}

\noindent where $[\hat{p},\hat{q}]$ is the \emph{commutator} of $\hat{p}$ and $\hat{q}$. The condition for the conservation of energy given in the text by \eqref{eq:energy-conservation} then becomes

\begin{align}
\frac{d}{dt}\mathrm{Tr}[\tilde{\rho}\tilde{H}] &= 0. \label{eq:energy-conservation-rho}
\end{align}

\noindent Using \eqref{eq:expectation-rho} and \eqref{eq:rho-time-dependence}, this is

\begin{align}
\frac{d}{dt}\mathrm{Tr}[\tilde{\rho}\tilde{H}] &= \mathrm{Tr}\left[(\tilde{\Omega}\tilde{\rho} - \tilde{\rho}\tilde{\Omega})\tilde{H}\right], \nonumber \\
&= \sum_{i}\left(\langle\phi_{i}|\tilde{\Omega}|\Psi\rangle\langle\Psi|\tilde{H}|\phi_{i}\rangle - \langle\phi_{i}|\Psi\rangle\langle\Psi|\tilde{\Omega}\tilde{H}|\phi_{i}\rangle\right), \nonumber \\
&= \sum_{i}\left(\langle\Psi|\tilde{H}|\phi_{i}\rangle\langle\phi_{i}|\tilde{\Omega}|\Psi\rangle - \langle\Psi|\tilde{\Omega}\tilde{H}|\phi_{i}\rangle\langle\phi_{i}|\Psi\rangle\right), \nonumber \\
&= \langle\Psi|(\tilde{H}\tilde{\Omega} - \tilde{\Omega}\tilde{H})|\Psi\rangle = 0. \nonumber
\end{align}

\noindent Again, in order to hold for arbitrary  $|\Psi\rangle$, we must have

\begin{align}
\tilde{H}\tilde{\Omega} - \tilde{\Omega}\tilde{H} &= 0, \nonumber
\end{align}

\noindent as given earlier in \eqref{eq:HOmega-condition}.

\end{document}